# 3D-Printed Photocathodes for Resonant, Terahertz-Field-Driven Ultrafast Electron Emission

Andrea Rovere[1], Riccardo Piccoli[1], Andrea Bertoncini[2], Young-Gyun Jeong[1], Stéphane Payeur[1], François Vidal[1], Seung-Heon Lee[3], Jin-Hong Seok[3], O-Pil Kwon[3], Roberto Morandotti[1,4], Carlo Liberale[2]*, Luca Razzari[1]**

[1]     INRS - Énergie, Matériaux et Télécommunications, Varennes, Québec J3X 1S2, Canada
[2]     KAUST, BESE (Biological Environmental Science and Engineering) division, Thuwal 23955-6900, Saudi Arabia
[3]     Department of Molecular Science and Technology, Ajou University, Suwon 16499, Republic of Korea
[4]     Institute of Fundamental and Frontier Sciences, University of Electronic Science and Technology of China, Chengdu 610054, China

* carlo.liberale@kaust.edu.sa
** razzari@emt.inrs.ca

**Abstract**
Ultrashort photoemitted electron bunches can provide high electron currents within sub-picosecond timeframes, enabling time-resolved investigations of ultrafast physical processes with nanoscale resolution. Non-resonant conductive nanotips are typically employed to realize nanoscale photoelectron sources with high brightness. However, such emitters require complex non-scalable fabrication procedures featuring poor reproducibility. Planar resonant antennas fabricated via photolithography have been recently investigated, also because of their superior field enhancement properties. Nevertheless, the electron emission from these structures is parallel to the substrate plane, which limits their practical use as electron sources. In this work, we present an innovative out-of-plane, resonant nanoantenna design for field-driven photoemission enabled by high-resolution 3D printing. Numerical and experimental evidences demonstrate that gold-coated, terahertz resonant nanocones provide large local electric fields at their apex, automatically ensuring out-of-plane coherent electron emission and acceleration. We show that the resonant structures can be conveniently arranged in an array form, for a further significant electron extraction enhancement via a collective terahertz response. Remarkably, such collective behaviour can also be harvested to boost photoemission from an individual nano-source. Our approach opens the path for a new generation of photocathodes that can be reproducibly fabricated and designed at will, significantly relaxing the requirement for intense terahertz drivers.

**Main text:**

The development of ultrafast pulsed electron sources has enabled time-resolved characterizations of a variety of processes at the nanometer scale via ultrafast electron microscopy and diffraction[1,2] (including chemical reactions,[3,4] phase transitions,[5] and membrane mechanical drumming,[6] among others[7,8]). In these experimental techniques, ultrafast electron bunches are commonly generated by focusing an amplified intense femtosecond laser beam onto a properly chosen photocathode material. Consequently, the minimum attainable electron emission area is limited by standard optical diffraction constrains. In the last years, a novel promising geometry for electron photoemission featuring metallic [9-15] and silicon-based nanotips,[16] as well as carbon nano-tubes[17] has been extensively investigated. In such configuration, the enhancement of the electric field at the tip apex, typically associated with the so-called "lightning rod effect" (i.e., the confinement of the electric field on a conductive surface with a small radius of curvature)[18], provides a two-fold advantage since i) it relaxes the need of extremely high optical intensities for photoemission, and ii) it guarantees a sub-diffraction nanoscale-sized emission area. The high-brightness and localized photoemission from non-resonant nanotips have already been successfully employed to achieve simplified ultrafast electron microscopy setups with high spatial and temporal resolution.[19-22] In general, the nature of the photoemission process is determined by the Keldysh parameter $\gamma = \sqrt{\phi/2U_P}$, which relates the material work function $\phi$ to the ponderomotive force of the electrons $U_P = e^2 E^2 / 4 m_e \omega^2$, where $E$ and ω are the amplitude and the optical frequency of the driving electric field, while $m_e$ and $e$ are the mass and the charge of the electron, respectively. When $\gamma \gg 1$, electron emission is mostly driven by the photon energy (multi-photon absorption regime) while for $\gamma \ll 1$ emission is predominantly dictated by the amplitude of the electric field (tunneling regime). Importantly, in the field-driven regime, the electrons are extracted within half optical cycle, thus ensuring a coherent emission of the electron bunches (i.e., in phase with the driving electromagnetic pulse). On the other hand, the subsequent motion of the extracted electrons is described by the adiabatic parameter $\delta = l_F/l_q$, where $l_F$ is the decay length of the optical field in proximity of the emitter surface, and $l_q = eE/m\omega^2$ is the amplitude of the electron quiver motion in the external electric field.[9] The use of low frequencies readily leads to the condition $\delta \ll 1$, which means that the electrons quiver amplitude becomes much larger than the near-field decay length at the emitter tip (see Figure S5a in SI), allowing the acceleration of the electrons in the enhanced near field region within the same half optical cycle of the incident electromagnetic pulse. It is therefore evident that the use of terahertz (THz) radiation (0.1-10 THz in frequency, 30-3000 μm in wavelength) is very appealing for the extraction and acceleration of electron bunches, ensuring ideal conditions for field-driven photoemission. Moreover, THz pulses generated via optical rectification (OR) are intrinsically carrier-envelope phase stable,[23] thus offering precise coherent control of the extracted electron bunches.[24] Recently, the field enhancement of *non-resonant* metal nanotips has been investigated for THz-driven photoemission,[25,26] to achieve nanoscale-localized electron extraction together with local THz electric field amplitudes of few MV/cm, which are required to observe a sizeable photocurrent. Concomitantly, other studies have exploited the higher field enhancement granted by planar *resonant* antennas with respect to non-resonant nanotips for a variety of driving optical frequencies, enabling electron photoemission at lower incident field amplitudes.[27-31] However, the close proximity of the substrate underneath the planar antennas hampers the practical use of the emitted electron bunches for the most relevant applications. In this work, we present innovative photocathode designs for out-of-plane THz-driven photoemission, enabled by the high reproducibility and precise control offered by advanced 3D printing lithography.[32] We exploit the strong monopolar resonance of gold vertical nanocones (resonant nanocones: RNCs), prepared on a conductive gold surface (Figure 1a) acting as a mirror-image plane,[33] to achieve a significant boost in field-driven electron emission in comparison with standard non-resonant nanotips. Moreover, we show that the collective response of an RNC array can be finely tuned to achieve an even higher local THz field at the RNC apexes, further enhancing the photoemission. Finally, taking advantage of the geometrical freedom granted by 3D printing fabrication, we demonstrate that this "cooperative effect" can also be engineered to confine the field-driven photoemission to an individual RNC emitter. This isolated emission is achieved by positioning a single emitting RNC in an array of resonant gold cylinders.

Different types of nanostructures and arrangements have been investigated: i) a single non-resonant nanotip (Figure 1b), ii) a single RNC (Figure 1c), iii) an array of RNCs (Figure 1d), iv) an array of "cooperative" RNCs (Figure 1e), and v) a single RNC in an array of cooperative cylinders (Figure 1f). In the cases of the resonant structures, we have tailored the THz response of the samples by means of numerical simulations (COMSOL Multiphysics software) centering the monopole resonance at around 1 THz. The samples have then been fabricated via a 3D printing technique based on two-photon absorption in a photopolymer (see SI).[32] The fabricated RNCs feature an aspect ratio of 1/10 (base diameter/height) and an apex radius of curvature of 150 nm, determined by the maximum resolution of the fabrication process. Subsequently, the entire array area, measuring $1.8 \times 1.8$ mm$^2$, has been covered by the sputtering of a 200-nm-thick gold layer, thus coating the fabricated RNCs as well as the bottom plane. To compare the photoemission characteristics of a commonly-used non-resonant metal nanotip with the ones of our resonant nanostructures, we have also fabricated a 1-mm-high gold-coated nanotip (*non-resonant nanotip*, Figure 1b), presenting an apex radius of curvature equal to the one of the RNCs. The field enhancement of a non-resonant metal nanotip can be analytically evaluated considering a hemisphere-capped cylindrical rod of radius $R$.[34] In this case, the field enhancement is given by $F = \alpha(\lambda/R)$, with $\alpha = 0.06$ [25] (Figure 2a). Thus, for our structure, $F = 120$ at 1 THz. The case of a single gold RNC on an infinite conductive surface (*single RNC*, Figure 1c) has been considered via numerical simulations, in which a p-polarized plane wave illuminates the RNC with an angle $\theta = 30°$ with respect to its principal axis (defined as the z axis hereon). It is worth noticing that only the z-component of the THz electric field ($E_z = E_0 \sin \theta$) contributes to excite the monopolar resonance of the RNC. In this case, the field enhancement is defined as $F = E_{loc}/E_0$, where $E_{loc}$ is the local electric field calculated at 1 nm from the apex of the RNC and $E_0$ is the incident electric field. The calculated maximum value of $F$ is about 900 at the resonance frequency, when the height of the RNC is set at approximately $\lambda_{res}/4 \approx 80$ µm (Figure 2b). The numerically calculated extinction cross-section of the single RNC also presents a monopolar resonance centered at around 1 THz, with a bandwidth (full-width at half maximum - FWHM) of 250 GHz. Note that the THz response of the single RNC cannot be directly measured via far-field THz spectroscopy due to the very low extinction cross-section associated with an individual nanostructure. The response of a periodic array of THz RNCs (*array of RNCs*, Figure 1d) has then been investigated. This design allows obtaining a coherent emission of electrons from multiple RNCs, for a spatially-structured electron pattern that also leads to a higher overall current in comparison with a single emitter. [15,16] The fabricated RNCs, separated by 140 µm, are in this case 120-µm-high, to compensate for the array induced-shift of the resonance position, providing a field enhancement factor of about 1850 at around 1 THz. The measured THz reflectance (normalized to the gold substrate) is reported in Figure 2c (solid line) and shows a very good agreement with the numerical simulations (dashed lines), both in terms of the resonance central frequency and bandwidth (about 115 GHz FWHM in the array case). The collective electromagnetic response of an array of RNCs has been further optimized to exploit the synergic interplay between neighbouring elements via a proper harvesting of their re-scattered radiation (*array of cooperative RNCs*, Figure 1e). Indeed, under precise geometrical conditions,[35] it is possible to obtain constructive interference between the local electric field acting on each RNC and the scattered fields from the surrounding elements. This can also be explained in terms of the coupling between the monopolar resonance of an RNC and the so-called lattice resonance of the array.[35,36] In this sample, the fabricated RNCs are 81.6-µm-high and separated by 200 µm (see SI for details regarding the lattice resonance determination), which results in a significant boost in the local field enhancement, reaching a peak value of about 6800 with a narrower resonance bandwidth (~60 GHz FWHM from the reflectance measurement, see Figure 2d) with respect to the non-cooperative array of RNCs. Importantly, as previously mentioned, the benefit offered by this cooperative effect can also be preserved for the case of a single RNC surrounded by resonant cylinders (*single RNC in an array of cylinders*, Figure 1f). In this sample, the height of the central RNC and the periodicity of the array are the same as the ones of the array of cooperative RNCs, while the height of the cylinders is reduced down to 66 µm, to align their monopolar resonance to the one of the RNC. As it is possible to see from Figure 2e, the THz reflectance measurement on this sample presents features similar to the one of the array of cooperative RNCs. The electric field enhancement value in proximity of the RNC apex is in this case assumed to be the same as the one extracted for the case of the array of cooperative RNCs, as suggested by numerical simulations conducted on a simplified model of 1 RNC surrounded by 8 cylinders (see Figure S1 in SI). As can be seen from Figure 2e, the maximum field enhancement at the RNC apex is 11 times higher than the one reached on the cylinders. Hence, since field-driven photoemission is highly nonlinear with respect to the THz local electric field, such a hybrid geometry guarantees a single "hot spot" for electron emission while preserving the benefits of a collective excitation.

In order to have access to a simple experimental evaluation of the THz-field-induced electron emission characteristics of the fabricated samples, we have exploited the fluorescence signal emitted by argon gas surrounding the metallic nanostructures under a controlled static pressure, in a manner similar to what has been introduced by K. Iwaszczuk *et al.*[28] Indeed, the gas atoms can be excited via the impact with the extracted electrons, resulting in a fluorescence signal that can be used to characterize the photoemission process through the Fowler-Nordheim (FN) relation[28] (see Equation 2 below). The illustration of the experimental setup used for this measurement is shown in Figure 3a. Broadband (0.1 – 3.0 THz) THz pulses with a maximum peak electric field of 180 kV/cm have been generated via OR in a 2-(4-hydroxy-3-methoxystyryl)-1-methylquinolinium 2,4,6-trimethylbenzenesulfonate (HMQ-TMS) nonlinear optical organic crystal pumped by an amplified Yb laser system emitting 1 mJ, 170-fs-long pulses.[37-38] In the experiments, two wire-grid polarizers have been used in series on the beam path to vary the incident THz field amplitude while maintaining a p-polarization state at their output. The sample has been placed in a sealed chamber filled with argon gas at a static pressure of 2.3 bar (for the pressure dependence of the fluorescence signal, see Figure S3b in SI). The THz beam has been sent through a z-cut quartz window onto the sample with an incident angle of 30 degrees with respect to the principal axis of the RNCs. The fluorescence intensity as a function of the input THz peak electric field amplitude has been acquired by means of a CCD camera, allowing a relative evaluation of the local fluorescence for each individual nanostructure in a sample (see Figure S4 in SI).

It is important to note that the sharp resonance of the metallic RNCs acts as a frequency bandpass filter for the impinging broadband THz radiation. Therefore, the observed photoemission process is the result of a trade-off between the absolute field enhancement peak value at the resonance frequency and the spectral portion of the THz excitation pulses effectively filling the RNC resonance bandwidth. For this reason, the key reference parameter becomes the time-domain THz peak electric field enhancement:

$$\beta_{calc} = \max_t(E_{local}) / \max_t(E_{exc}) \qquad (1)$$

where $E_{local}$ is the value of the local electric field evaluated 1 nm above the nanostructure apex and $E_{exc}$ is the effective incident electric field on the sample (i.e., considering the transmission of the sample chamber window and the electric field projection on the axis of the RNC; $\max_t(E_{exc}) = 80$ kV/cm for the maximum input peak electric field in our experiments). $\beta_{calc}$ thus implicitly takes into account the spectral overlap between the incident THz pulses and the samples resonance (see detailed explanation and Figure S5b in SI). In the field-driven regime, the nonlinear relation between the tunneling current density $J$ (assumed to be directly proportional to the recorded fluorescence intensity) and the impinging THz peak electric field $E_{exc}$ is described by the FN equation:[39]

$$J = a \frac{(\beta E_{exc})^2}{\phi} exp\left(-b \frac{\phi^{3/2}}{\beta E_{exc}}\right) \qquad (2)$$

where $a = 1.54 \times 10^{-6} \; A \; eV \; V^{-2}$ and $b = 6.83 \times 10^9 \; eV^{-3/2} \; V \; m$ are the FN constants. In order to make an appropriate comparison between the tested nanostructures, the fluorescence intensity has been evaluated for an individual emitter per each sample. It is convenient to rewrite Equation 2 as:

$$\log\left(J/E_{exc}^2\right) = C_1 + C_2/E_{exc} \qquad (3)$$

where $C_1 = \log\left(\frac{\phi}{a\beta^2}\right)$ and $C_2 = -\frac{b\phi^{3/2}}{\beta}$. In this way, it is possible to represent the experimental data in a linear plot. All the measurements exhibit a high degree of linearity, as it can be seen in Figure 3b-f, confirming the field-driven nature of the photoemission process.[17] Through this representation, via a simple linear fit we can extract the slope factor $C_2$ for each structure, which is directly related to the time-domain field enhancement factor, $\beta_{exp}$:

$$\beta_{exp} = -\frac{b\phi^{3/2}}{C_2} \qquad (4)$$

the values of $\beta_{exp}$ retrieved by fitting the data from the fluorescence experiments using Equation 4 represent an effective way to compare the photoemission performance of the different nanostructures. In this comparison, we have considered an effective work function of $\phi = 0.78 \pm 0.03$ eV for all the gold nanostructures. Such value has been retrieved by fitting the experimental data for the non-resonant nanotip while fixing $\beta = 128$ since, as explained above, a well-established analytical formula exists for this case.[34] As already observed in ref.[28] for evaporated gold antennas, the obtained value is lower than the work function typically reported for bulk gold (~5 eV).[40] This difference has been attributed to i) impurities and adsorbates from air deposited on the photoemission area, which can locally lower $\phi$ of a metal under field-driven electron tunneling;[41] ii) the intrinsic roughness of the evaporated gold surface, with nanometer-scale features at the nanostructure apex, which may lead to an increased local field enhancement. Indeed, an underestimation of $\beta$ would result in a decreased retrieved value for $\phi$, due to the dependence of the slope factor in Equation 4 on both these two parameters. Table 1 summarizes the time-domain field enhancement values $\beta_{exp}$ retrieved from the fluorescence experiments. They are in good agreement with the corresponding enhancement factors $\beta_{calc}$ returned by the numerical simulations (Equation 1). As one can see, the $\beta$ values relative to the RNCs are higher than the one of the non-resonant nanotip, indicating a more efficient electron photoemission. Additionally, the exploitation of the cooperative effect in an array geometry leads to a final ~2.5 times increase in the field enhancement factor with respect to the non-resonant case, confirming a similar photoemission boost for the case of the array of cooperative RNCs and the single RNC in the array of cooperative cylinders. The time-domain field enhancement value not only drives the tunneling process but is also the key quantity in determining the subsequent electron acceleration and thus the resulting electron energy spectrum, an important parameter for practical applications. The overall process can be described via a two-step analytical model that allows to retrieve the kinetic energy of the accelerated electrons.[17] According to this model, the cooperative effect is predicted to result in an almost two-fold increase in the electron energy cut-off with respect to the case of a non-resonant nanotip, with values extending beyond 400 eV for just 80 kV/cm of effective incident electric field $E_{exc}$ (see section 6 of SI for calculation details).

Finally, proof-of-principle electrical measurements have also been performed by means of an in-house built setup, to provide direct experimental evidence of electron emission. To this end, a solid-state electrometer (Keithley 610C) has been employed. The sample under investigation has been placed in a chamber under ultra-high vacuum (~$10^{-6}$ mbar) and a copper foil anode (2 x 2 cm²) connected to the electrometer has been used to collect the emitted electrons (Figure 4a). The measured photocurrents produced by the array of RNCs and the array of cooperative RNCs (Figure 4b) as a function of $E_{exc}$ present the expected threshold-like behavior, similar to the one observed in the fluorescence experiments, thus further confirming the field-driven process. The current starts to be observable at an excitation THz peak electric field of ~50 kV/cm and reaches a maximum at 80 kV/cm of about 1.5 pA for the array of RNCs and 1.15 pA for the cooperative array, corresponding to 6 fC and 4.6 fC per pulse ($3.75 \times 10^4$ and $2.87 \times 10^4$ electrons per pulse), respectively. It is important to underline that the extracted electron bunches in our setup experience a considerable transversal spread, due to space charge effects (up to several cm, see SI for numerical simulations estimates), during the propagation from the photocathode to the anode (1.5 cm distance in our setup, also determined by the necessity of not blocking the incoming THz radiation). Thus, the geometrical constraints of our basic detection scheme hinder the complete collection of the total photoemitted current. This likely affects the measurement of the array of cooperative RNCs even more severely than the one of the non-cooperative array, considering that the RNCs of the former sample are endowed with a larger $\beta$ value and are then expected to deliver a higher number of electrons per THz pulse (see SI for details). Furthermore, the two RNC arrays feature a different number of RNCs per unit area (~2 times smaller in the case of the array of cooperative RNCs), which also affects the overall measured current in the two cases. These facts prevent the use of such electrical characterization for a quantitative comparison between the arrays. Note that, for the samples with a single emitting nanostructure, the photocurrent measured using our in-house developed setup resulted to be too weak to return consistent datasets.

To test the robustness of our nanostructures, the emitted photocurrent has then been acquired for long exposure times and under the highest incident THz field provided by our source. As we can see in the inset of Figure 4b, the gold RNCs are still functional after several hours of continuous THz illumination. The presence of a slow photocurrent decay in time, beside a certain degree of possible degradation of the emitter surface, is also related to a residual charge accumulation on the anode that was observed in our setup in such long measurements. The structural stability of the RNCs has been further confirmed by an SEM inspection of the nanostructures after the prolonged THz exposure of our tests (> $10^7$ THz pulses). As it is possible to observe in Figure 4c, the RNC gold surface does not show signs of critical damage, especially in proximity of the RNC apex, which is typically the most solicited area in these experiments.[25]

In conclusion, we have proposed a novel photocathode design for ultrafast field-driven photoemission exploiting the advantages of high-resolution 3D printing. Our approach allows achieving out-of-plane electron emission and acceleration employing broadband table-top THz sources delivering peak electric fields at the 100 kV/cm level. By means of numerical modelling, THz far-field characterization, and the analysis of electron-induced argon gas fluorescence, we have demonstrated the advantages offered by nanostructures featuring a monopolar resonance at THz frequencies with respect to traditional non-resonant nanotips. In particular, by tailoring the collective response of RNCs arranged in an array geometry, a significant boost of the nano-localized THz peak electric field of a factor of 2.5 when compared to the standard non-resonant case has been obtained in our experimental conditions (meaning that the peak electric field of the THz driver can be equivalently reduced by the same factor to achieve the same level of electron extraction/acceleration). Remarkably, such boost can also be harvested when electron emission from an individual nanostructure is targeted, as shown for the case of an RNC in an array of cylinders. We have also performed proof-of-principle photocurrent measurements that have corroborated the fluorescence intensity characterization, confirming the field-driven nature of the electron emission. A maximum current of 1.5 pA has been recorded, corresponding to $3.75 \times 10^4$ electrons per THz pulse (likely an underestimate of the actual extraction, due to the limits of our in-house built setup), showing that the developed scheme can deliver high brightness electron bunches of interest for, e.g., time-resolved electron diffraction experiments. Finally, the fabricated nanostructures have shown a good structural resistance after a prolonged exposure to intense THz pulses (more than $10^7$ THz shots). While the effectiveness of our strategy has been demonstrated for the case of broadband THz sources, it is worth mentioning that the use of a narrowband source would grant access to the full exploitation of the sharp resonance of the investigated nanostructures. Since narrowband THz sources at the kV/cm peak electric field level can be built with femtosecond lasers of a few μJ energy[42], this configuration could open the path to a novel class of THz-driven ultrafast electron sources featuring unprecedentedly high (MHz-level) repetition rates. We thus envision that the combination of the most recent advances in THz source technology with rationally-designed 3D-printed resonant nanostructures may find important applications in the development of next-generation nano-localized and coherent ultrafast electron sources.

**Acknowledgements**
This work was supported by the Natural Sciences and Engineering Research Council of Canada (NSERC) through the Strategic and Discovery grant programs as well as by the Canada Research Chair program. O.-P. Kwon would also like to acknowledge support from the National Research Foundation of Korea (NRF) funded by the Ministry of Science, ICT & Future Planning, Korea (No. 2014R1A5A1009799). R. M. is affiliated to the Institute of Fundamental and Frontier Sciences, University of Electronic Science and Technology of China as an adjoint faculty.

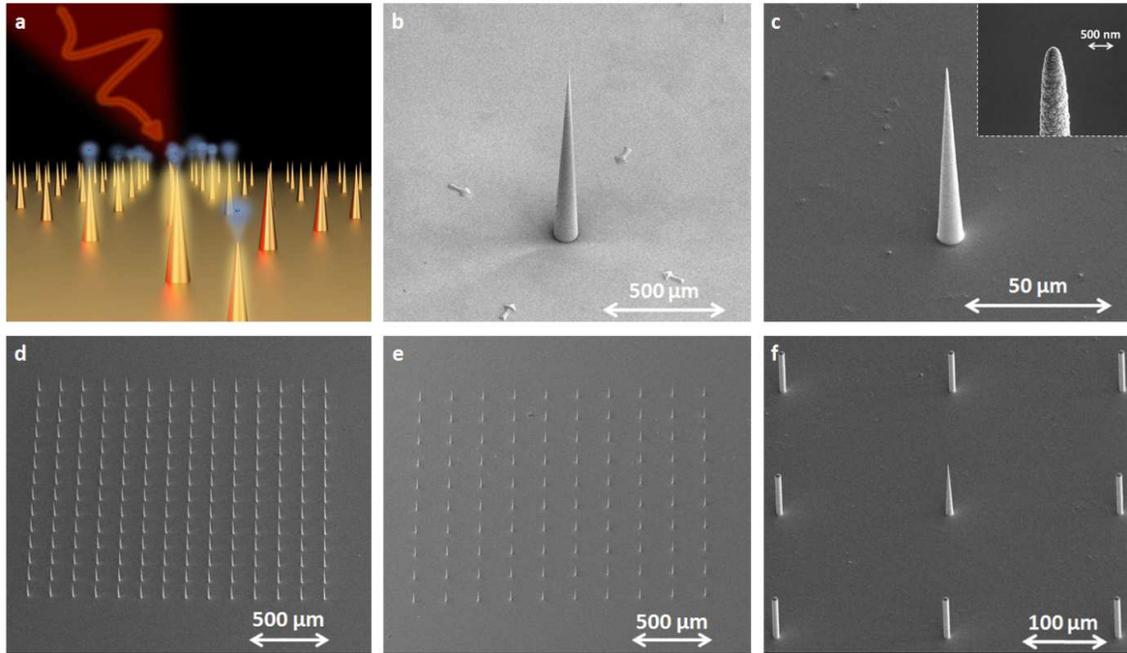

**Figure 1.** a) Illustration of THz field-driven electron photoemission from an array of monopolar gold RNCs. SEM pictures of the fabricated samples: b) the 1-mm-high non-resonant nanotip, c) the single RNC, d) the array of RNCs, e) the array of cooperative RNCs, and f) the single RNC in an array of cylinders. The apex of an RNC is shown in the inset of Figure 1c.

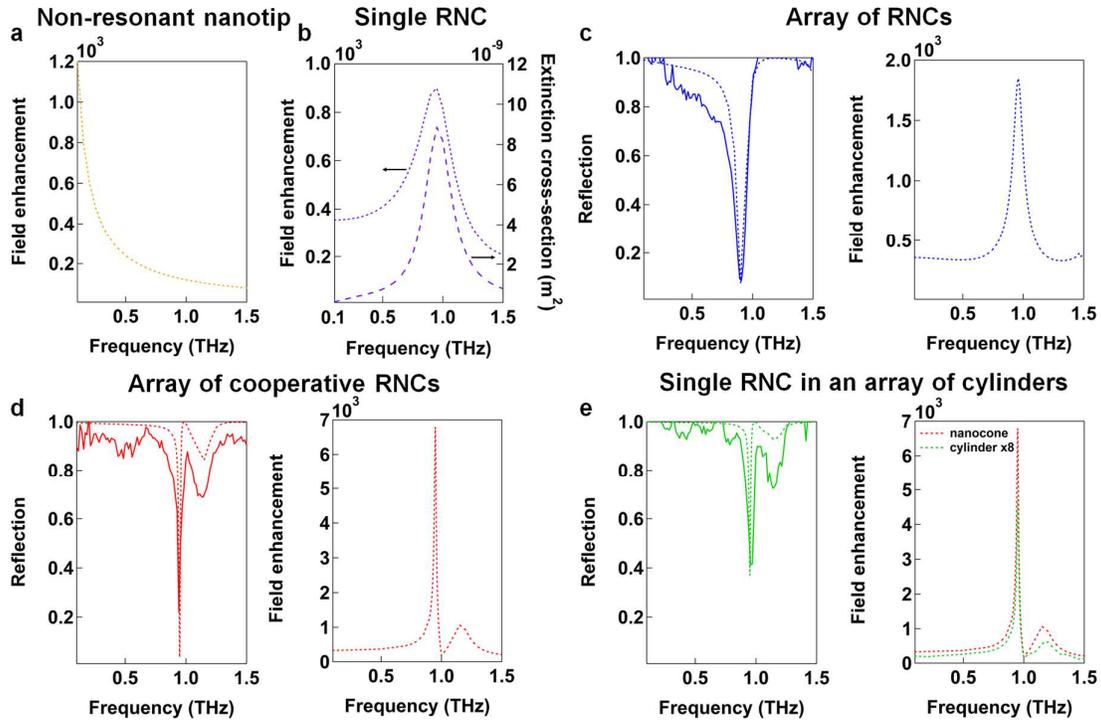

**Figure 2.** THz frequency response of the investigated samples. The illustrations display the near- and far-field properties of the samples, as numerically calculated via COMSOL Multiphysics, or analytically retrieved for the case of the non-resonant nanotip field enhancement (dotted lines), as well as the measured reflectance of the arrays (solid lines). a) Non-resonant nanotip, b) Single RNC, c) Array of RNCs d) Array of cooperative RNCs and e) Single RNC in an array of cylinders. In the latter case, we simulated an array of cylinders with a radius of curvature at their edges equivalent to the one of the apex of the RNCs. As it is shown, the field enhancement calculated in proximity of the top of the cylinders is about 11 times lower than the one of the central RNC, which is considered to be comparable with the one retrieved for the case of the array of cooperative RNCs (see SI for further details).

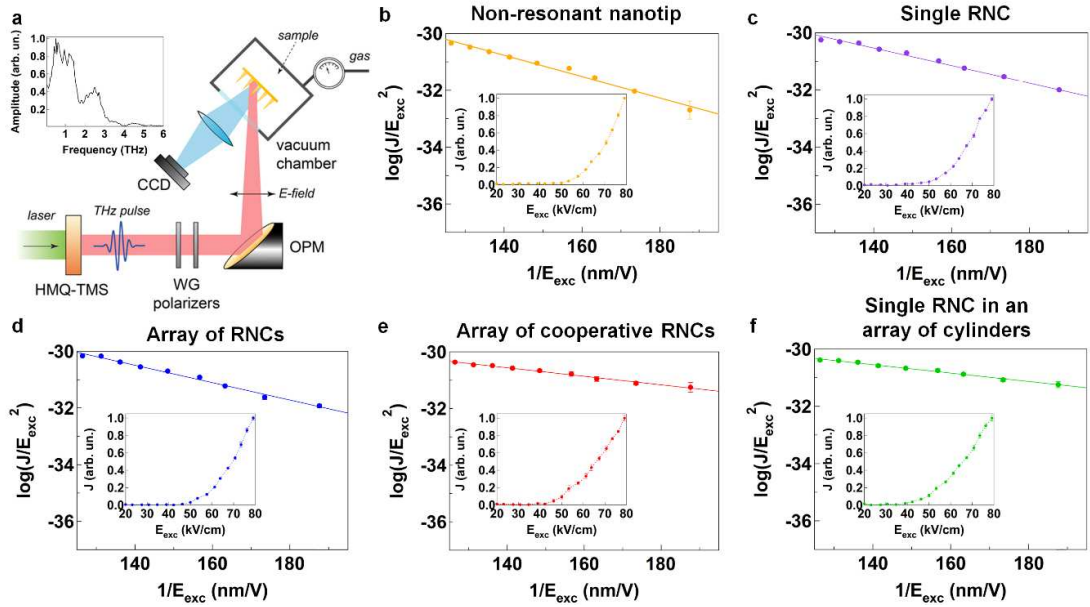

**Figure 3.** Fluorescence measurements. a) Schematic of the setup used for the fluorescence measurements. WGP – wire grid polarizers; OPM – 90 deg off-axis parabolic mirror. The spectrum of the incident THz pulse is reported in the inset. b-f) Experimental data (dots) of the measured argon fluorescence (see SI for details) and linear fit using Equation 3 (solid lines) for the various samples. Insets show the raw measured fluorescence values in a linear scale, as a function of the incident electric field.

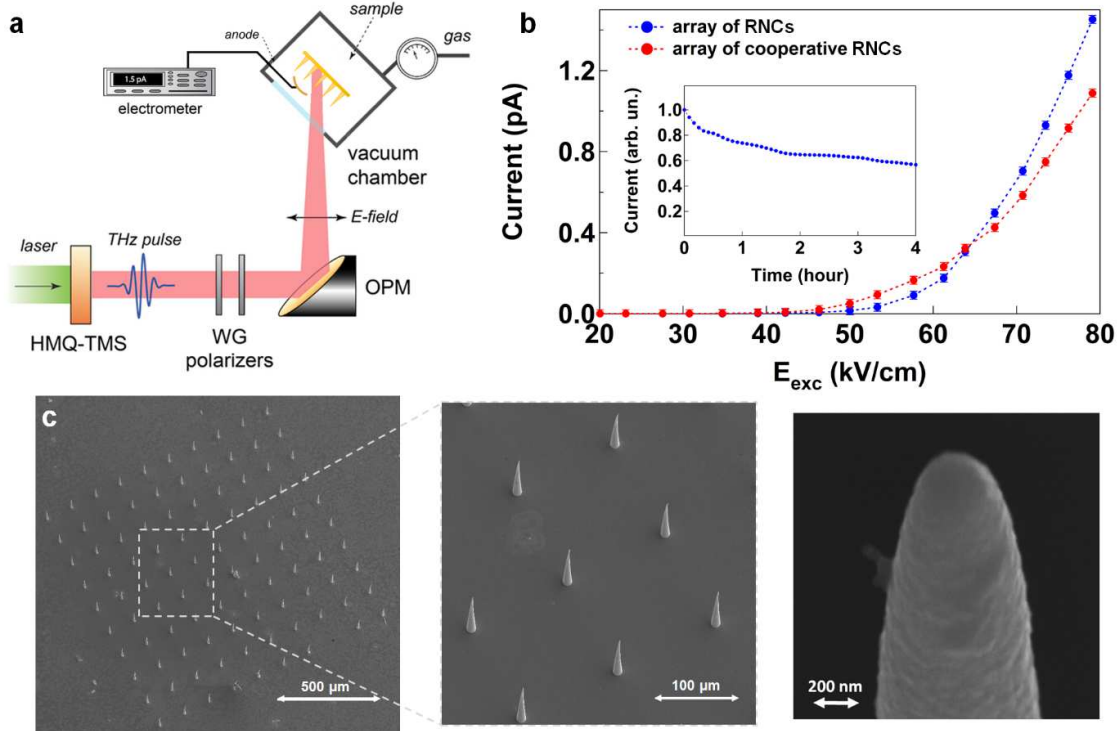

**Figure 4.** Photocurrent measurements. a) Schematic of the experimental setup used to measure the photocurrent. b) Current as a function of the incident THz peak electric field for the two RNC array samples. The stability of the emission for the array of RNCs under a 4-hour-long exposure to the highest THz field is reported in the inset. c) SEM images of the array of RNCs taken after the characterization experiments.

**Table 1.** Slope factors $C_2$ for each sample, retrieved from the fit of the fluorescence experiments presented in Figure 3, using Equation 3. The corresponding time-domain field enhancement values extracted from these measurements ($\beta_{exp}$) and the calculations ($\beta_{calc}$). The case of the non-resonant nanotip has been used as a reference to estimate the gold work function ($\phi = 0.78 \pm 0.03$ eV), by fixing $\beta$ to the value retrieved analytically.

| Parameter | Non-resonant nanotip | Single RNC | Array of RNCs | Array of cooperative RNCs | RNC in an array of cylinders |
|---|---|---|---|---|---|
| $C_2$ | $(-3.7 \pm 0.2) \times 10^7$ | $(-3.1 \pm 0.1) \times 10^7$ | $(-3.1 \pm 0.1) \times 10^7$ | $(-1.51 \pm 0.06) \times 10^7$ | $(-1.47 \pm 0.05) \times 10^7$ |
| $\beta_{exp}$ | 128 (analytical) | $153 \pm 10$ | $153 \pm 10$ | $314 \pm 21$ | $322 \pm 21$ |
| $\beta_{calc}$ | 128 (analytical) | 204 | 209 | 344 | 344 |